\documentstyle[12pt]{article}\textheight= 22.8cm\textwidth=
15.65cm \columnsep=0.5cm\voffset= -2cm\hoffset= -1.4cm
\begin{document}\def\p{\phi}\def\P{\Phi}\def\a{\alpha}\def\e{\epsilon}
\def\be{\begin{equation}}\def\ee{\end{equation}}\def\l{\label}
\def\0{\setcounter{equation}{0}}\def\b{\beta}\def\S{\Sigma}\def\C{\cite}
\def\r{\ref}\def\ba{\begin{eqnarray}}\def\ea{\end{eqnarray}}
\def\n{\nonumber}\def\R{\rho}\def\X{\Xi}\def\x{\xi}\def\la{\lambda}
\def\d{\delta}\def\s{\sigma}\def\f{\frac}\def\D{\Delta}\def\pa{\partial}
\def\Th{\Theta}\def\o{\omega}\def\O{\Omega}\def\th{\theta}\def\ga{\gamma}
\def\Ga{\Gamma}\def\t{\times}\def\h{\hat}\def\rar{\rightarrow}
\def\vp{\varphi}\def\inf{\infty}\def\le{\left}\def\ri{\right}
\def\foot{\footnote}\def\ve{\varepsilon}\def\N{\bar{n}(s)}\def\cS{{\cal S}}
\def\k{\kappa}\def\sq{\sqrt{s}}\def\bx{{\bf x}}\def\La{\Lambda}
\def\bb{{\bf b}}\def\bq{{\bf q}}\def\cp{{\cal P}}\def\tg{\tilde{g}}
\def\cf{{\cal F}}\def\bN{\hat{N}}\def\Re{{\rm Re}}\def\Im{{\rm Im}}
\def\K{\hat{K}}\def\cH{{\cal H}}\def\vs{\varsigma}
\large

\title{Quantization of Solitons in Coset Space}
\author{ J.Manjavidze$^{a)}$, A.Sissakian$^{b)}$}
\maketitle

\vskip 0.8cm
\begin{abstract}
The perturbation theory around the soliton fields of the sin-Gordon
model is developed in the coset space. It is shown by explicit
calculations that all corrections to the topological soliton
contribution are canceled exactly.
\end{abstract}

PACS numbers: 02.30.Cj, 03.65.Db, 02.40.Vh, 31.15.Kb
\newpage
{\large\bf I. Introduction}
\vskip 0.3cm

The problem of quantization of the extended objects was formulated
mainly in the middle of 70-th, see the review paper $^1$ and
references cited therein. One starts from the classical Lagrange
equation:
\be
\f{\d S(u)}{\d u(x,t)}=0,
\l{0.1}\ee
where, for simplicity, $u(x,t)$ is the real scalar field $^{2}$.  If
this equation has nontrivial solution $u_c(x,t)$ then the problem of
its quantization will arise. One of the first attempts to construct
the perturbation theory was based on the WKB expansion  in vicinity
of $u_c$ $^{3}$.

The Born-Oppenheimer method was adopted also $^{4,5}$. First of all,
to construct the quantum mechanics the structure of Hilbert space
$\cH$ is postulated.  So, it is assumed that the Fock column consist
from the vacuum state $|0>$ and from the multiple meson states
$|p_1,p_2, ..., p_n>$, $n\geq1$.  The ordinary perturbation theory
operates just with this `meson' sector only. The $anzats$ $|P_1,
P_2,...,P_l>$ $^5$ for the $l$-soliton state, $l\geq 1$, is introduce
also.

It is postulated that the quantum excitations in the soliton sector
are described by the excitation of the `meson' field $^4$.
Therefore, to construct the perturbation theory, there should also be
the mixed states:
\be
|P_1,...,P_l;p_1,...,p_n>,~l\geq1,~n\geq1,
\l{0.2}\ee
but, at the same time,
\be
<P_1,...,P_l;p_1,...,p_n|p_1,...,p_{n'}>\equiv0,~l\geq1,~n+n'\geq0,
\l{0.3}\ee
i.e. it is assumed that the solitons are the absolutely stable field
configurations $^1$.

Present paper in definite sense completes the offered in $^{4,5}$
picture.  The (1+1)-dimensional exactly integrable sin-Gordon model
will be considered to illustrate our result. We will investigate the
multiple production of `mesons' by `soliton' and the truth of
(\r{0.3}) will be shown at the end of explicit calculations. In
other words, it will be shown that the offered in $^{4,5}$ postulate
concerning orthogonality of the `meson' $\cH_m$ and `soliton' $\cH_s$
Hilbert spaces can be proved. We will see that this conclusion
follows from exactness of the semiclassical approximation for
sin-Gordon model.

It should be noted that the exactness of semiclassical approximation
in topological soliton sector of the sin-Gordon model is not `beyond
the realm' $^6$. It is well known also that the integrable Coulomb
problem is exactly semiclassical.  The same we have for the quantum
rigid rotator $^{7}$, which is the isomorphic to Poshle-Teller model.
The general discussion of the exactness of semiclassical
approximation from a geometrical point of view was given in $^{8}$.

It will be crucial for us in many respects to follow the WKB
ideology. So, we will consider the `meson' production amplitudes
\be
a_{nm}(p,q)=<p_1,...,p_n|q_1,...,q_m>_{s},~n,m=1,2,...
\l{0.5}\ee
The index $s$ means that the calculations are performed in the
`soliton' sector and $p_i$ and $q_i$ are the `meson' momenta.  By
definition,
\be
p_i^2=q_i^2=m^2
\l{0.5a}\ee
since the quantum uncertainty principle leads to the impossibility
off mass-shell observation of the field $^{9}$. The ordinary
reduction formalism will be used to calculate $a_{nm}$. This means
that we will construct the $phenomenological$ $S$-matrix of the
`meson' interaction through the `soliton' fields, i.e. we will start
from the assumption that the states (\r{0.2}) exist, and it will be
shown at the end of calculations that such $S$-matrix is trivial:
\be
a_{nm}(p,q)\equiv0,~n+m>0.
\l{0.6}\ee

Offered in the paper formalism allows to prove (\r{0.6}). For this
purpose we will build the perturbation theory expansion over $1/g$,
where $g$ is the interaction constant $^{10}$. This perturbation
theory is dual to the theory described in $^{1}$, over $g$, i.e. one
can not decompose the definite order over $g$ contribution in terms
of the $1/g$ expansion, and vice versa. So, only the summary results
of both expansion may be compared.

Following to WKB ideology, to find the corrections to the
semiclassical approximation in the vicinity of the extremum
$u_c(x,t)$, one should find the solution of the equation for the
Green function:
$$
(\pa^2 + v''(u_c))G(x,t;x',t')=\d(x-x')\d(t-t'),
$$
where $v''(u)$ is the second derivative of the potential function
$v(u)$.  This Green function describes propagation of a particle in
the time dependent inhomogeneous and anisotropic external field
$u_c(x,t)$.  Generally, this problem has not a closed solution. So,
for instance, the attempt to solve the problem using the momentum
decomposition $^{11}$ leads to the hardly handling double-parametric
perturbation theory. To avoid this problem we will build new
perturbation theory over $1/g$.

Imagining particles coordinates as the elements of the Lee group, the
classical particles motion may be described mapping the trajectory on
group manifold. Roughly speaking, this means that the group
combination law creates the particles classical trajectory $^{12}$.

Moreover, this program was realized for description of the particle
quantum motion $^{13}$. It was shown for essentially nonlinear
Lagrangian $ L=\frac{1}{2}g_{\mu \nu}(x)\dot{x}^{\mu}\dot{x}^{\nu} $
that the semiclassical approximation is exact on the (semi)simple Lee
group manifold. But this slender solution of quantum problems is
destructed in presence of the interaction potential $v(x)=O(x^n)$,
$n>2$, since last one breaks the isotropy and homogeneity of the Lee
group manifolds $^{10}$. The developed perturbation theory will
describe the quantum perturbations breaking isotropy and homogeneity
of the group manifold.

Developed formalism contains the following steps $^{10,14}$. (i) We
will introduce the manifold $W_G$ of trajectories $u_c$, solving the
eq.(\r{0.1}).  The manifold $W_G$ will be labeled by the local
coordinates $(\x,\eta)$, i.e. we will consider $u_c=u_c(x;\x,\eta)$
since $u_c$ should belong to $W_G$ $completely$.  (ii) The numbers
$(\x,\eta)$ are interpreted as the generalized coordinates of the
`particle'. Then $u_c(x;\x,\eta)$ will define the external potential
for it.  The quantum motion of the `particle' may be described
noting that $W_G$ is the homogeneous and isotropic manifold, since
this case is rather quantum mechanical problem in the `flat' space.

It was shown in $^{14}$ that the WKB model $^3$, where the field
excitations in vicinity of $u_c$ are decomposed over the `meson'
states, and our model quantum mechanics of the `particle' in the
external potential defined by $u_c$, are isomorphic. In other words,
we know that the quantum trajectory of the `particle' covers the
phase space $(\x,\eta)\in W_G$ densely.  But it should be noted also
that described in $^3$ model presents the expansion over the
interaction constant $g$ and our perturbation theory describes
expansion over the $(1/g)$.

In the classical limit (labeled by the index `$0$') the motion of our
`particle' must be free $^{14}$, i.e. its velocity should be a
constant,
\be
\dot\x_0=const, \dot\eta_0=0.
\l{0.7}\ee
This may be achieved expressing the set $\{\eta\}$ through the set of
generators of the subgroup broken by $u_c$ $^{15}$. It is evident,
such choice of the `particles' coordinate gives the same effect as in
the above discussed transformation to the homogeneous and isotropic
(semi)simple Lee group manifold $^{10}$, see also $^{16}$.  Moreover,
we will see that even in the case of nontrivial potential function,
one can get to the free `particles' motion, rescaling the quantum
sources $^{10,14}$.

Thus, the necessary invariant subspace $W_G$ would be chosen equal to
the coset space $G/G_c$:
\be
W_G=G/G_c,
\l{0.7a}\ee
where $G$ is the symmetry group and $G_c\subset G$ is the classical
solutions $u_c$ symmetry group. The problem of quantization of the
coset space have a reach history, see e.g. $^{17}$.  Described in
$^{10,14}$ formalism presents one of possible realization of the
coset spaces quantization scheme.

The last one means that we will realize the transformation generated
by the classical trajectory $^{14}$:
\be
u_c:~(u,p)(x,t)\to(\x,\eta)(t)
\l{0.8}\ee
Such construction of perturbation theory in the $W_G$ space require
the additional effort noting that the dimension of the original phase
space $(u,p)\in T^*V$ is infinite. Therefore, (\r{0.8}) assumes the
infinite reduction since the dimension of coset space $W_G$ is
finite $^{18}$. The crucial for us reduction scheme was formulated in
$^{14}$.

In other words, quantizing the sin-Gordon soliton fields, the space
coordinate would be an irrelevant variable. This is the well known
fact, e.g. $^3$, and it leads to the Lorentz non-covariant
perturbation theory. It is the consequence of absolute stability of
the solitary waves profile, i.e. of conservation of the topological
charge.  The necessary information concerning this question will be
given in Sec.3.

Having the complete theory, one can analyze the perturbations.
The crucial point of the new perturbation theory is the statement
$^{10}$ that the quantum corrections are accumulated strictly on
the boundaries $\pa W_G$ (bifurcation manifolds $^{19,20,15}$) of the
$W_G$ space.  Therefore, if
\be
\pa u_c\bigcap\pa W_G=\emptyset,
\l{0.10}\ee
then the problem is exactly semiclassical. On other hand, (\r{0.10})
means conservation of the topological charge: $\pa u_c$ is the flow
induced by the quantum perturbations in $W_G$ and if (\r{0.10}) is
not satisfied, then a flow into the forbidden domain with other
topological charge, separated by the bifurcation boundary, should
exist. So, (\r{0.10}) is the topological charge conservation law.

Notice, the solution (\r{0.10}) leads to (\r{0.6}) since particles
production is the pure quantum effect.  This will be shown in Sec.4.

The paper is organized as follows. In Sec.2 we will (i) formulate
the necessary for us boundary conditions to derive the LSZ reduction
formulae, (ii) find the explicit expression for $a_{nm}$, (iii)
formulate the mapping into the coset space $W_G$. In Sec.3
we will (i) consider the sin-Gordon model, (ii) discuss
necessary in the coset space boundary condition, (iii)
remind the structure of the new perturbation theory $^{14}$, (iv)
describe `meson' multiple production to show (\r{0.6}).

\vskip 0.5cm
{\large\bf II. Density matrix on the Dirac measure}
\vskip 0.3cm

Main point of this section is the attempt to generalize the ordinary
for field theory boundary condition:
$$
u(x\in\s_\infty)=0,
$$
where $\s_\infty$ is the remote hypersurface. This boundary condition
is used to remove the surface term, and it is necessary to formulate
the reduction formalism.  We would like introduce the new boundary
condition to have a possibility to include the non-vanishing on
$\s_\infty$ field configurations and, at the same time, throw off the
surface term.


The $(n+m)$-point Green functions $G_{nm}$ are introduced
through the generating functional $Z_{j}$ $^{21}$:
\be
G_{nm}(x,y)=(-i)^{n+m}\prod_{k=1}^{n}\hat{j}(x_k)\prod_{k=1}^{m}
\hat{j}(y_k)Z_j,
\l{2.1}
\ee
where $\hat{j}(x)={\d}/{\d j(x)}$ and the generating functional
\be
Z_j=\int Du e^{iS_j(u)}.
\l{2.3}\ee
The action
\be
S_j (u)=S(u)-V(u)+\int dx dt j(x,t)u(x,t),
\l{2.4}
\ee
where
\be
S(u)=\int dxdt\le(\f{1}{2}(\pa u)^2-m^2u^2\ri),~m^2\geq0,
\l{2.4'}\ee
is the free part and $V(u)$ describes the interactions.  At the end
of calculations one should put $j=0$.

To provide convergence, the integral (\r{2.3}) will be defined on the
Mills complex time contour $C_+$ $^{22}$.  For example,
\be
C_\pm :~t\to t+i\ve, \ve\to+0,~ -\infty\leq t\leq+\infty
\l{2.5}\ee
and after all calculations, one should return the time contour on the
real axis putting $\ve=0$.

In a `meson' sector the integration in (\r{2.3}) is performed over
all field configurations with standard vacuum boundary condition:
\be
\int d^2x \pa_{\mu}(u\pa^\mu u)=
\int_{\sigma_{\infty}}d\sigma_{\mu}u\pa^{\mu}u=0.
\l{2.6}\ee
It follows from this conditions that
\be
u(x\in\s_\infty)=0,~pa_\mu u(x\in\s_\infty)=0.
\l{2.6a}\ee
It excludes a contribution from the surface term since assumes that
field disappeared on the remote hypersurface $\s_\infty$.
Considering the `soliton' sector this boundary condition require the
modification since there is in the $(x-t)$ space such direction along
which the soliton field does not disappeared. The integral (\r{2.3})
would have a formal meaning till this boundary condition will not be
specified.

Let us introduce now the field $\vp$ through the equation:
\be
-\frac{\d S(\vp)}{\d\vp(x,t)} =j(x,t).
\l{2.7}\ee
It is assumed that we can formulate such boundary condition that the
surface term may be neglected calculating the variational derivative
in (\r{2.7}).  Then we perform the ordinary shift $u\to u+\vp$ in
integral (\ref{2.3}).  Considering $\vp$ as the probe field created
by the source:
\be
\vp(x)=\int d^2x' G_0(x-x')j(x'),~~
(\pa^2 +m^2)G_0 (x-x')=\d(x-x'),
\l{2.8}\ee
the connected Green function $G^{c}_{nm}$ only will be
interesting for us:
\be
G_{nm}^{c}(x,y)=(-i)^{n+m}\prod_{k=1}^{n}\hat{j}(x_k)
\prod_{k=1}^{m}\hat{j}(y_k)Z(\vp),
\l{2.9}\ee
where
\be
Z(\vp)=\int Du e^{iS(u)-iV(u+\vp)}
\l{2.10}\ee
is the new generating functional.

To calculate the nontrivial elements of $S$-matrix we must put the
external particles on the mass shell. Formally this procedure means
amputation of the external legs of $G^{c}_{nm}$ and further
multiplication on the free particles wave functions.  In result the
amplitude of $n$- into $m$-particles transition $a_{nm}$ in the
momentum representation has the form:
\be
a_{nm}(q,p)=(-i)^{n+m}\prod_{k=1}^{n}\hat{\vp}(q_k)
\prod_{k=1}^{m}\hat{\vp}^* (p_k) Z(\vp).
\l{2.11}\ee
Here the particles creation operator:
\be
\h{\p}^*(q)=\int d^2xe^{iqx}\h{\p}(x),~\h{\p}(x)=\frac{\d}{\d\p(x)}.
\l{2.12}\ee
was introduced. The eq.(\r{2.11}) is the ordinary LSZ reduction
formulae. But one should remember that the boundary condition
(\r{2.6}) should be generalized to have a permission for inclusion of
the soliton contributions calculating $Z(\vp)$.

Describing particles multiple production it is enough to consider the
generating functional:
$$
\R(\a,z)=
\exp\{-\int
d\O_1(p)\le(\h{\vp}^*_+(p)\h{\vp}_-(p)e^{i\a_+p}z_+(p)\ri.+
$$\be
+\le.\h{\vp}^*_-(p)\h{\vp}_+(p)e^{i\a_-p}z_-(p)\ri)\}
Z(\vp_+)Z^*(\vp_-),
\l{2.13}\ee
where
$$
d\O_n(p)=\prod_{k=1}^n\f{d^1p_k}{(2\pi)2\e(p_k)}=
\prod_{k=1}^nd\O_1(p_k), ~\e(p)=(p^2+m^2)^{1/2}.
$$
Let us calculate
$$ \le.
\int \f{d^2\a_+}{(2\pi)^2}e^{-iP\a_+} \f{d^2\a_-}{(2\pi)^2}e^{-iP\a_-}
\prod_{k=1}^n \f{\d}{\d z_+(p_k)}\prod_{k=1}^m \f{\d}{\d z_-(q_k)}
\R(\a,z)\ri|_{z_+=z_-=0}
$$
Inserting here the definition (\r{2.13}), one can find that this
expression gives:
$$
\d(P-\sum^n_{k=1}p_k)\d(P-\sum^m_{k=1}q_k)|a_{nm}(p,q)|^2,
$$
where the $\d$- functions are the result of integration over
$\a_\pm$. So, the factors $e^{i\a_\pm p}$ in (\r{2.13}) permit to
introduce the energy-momentum shell and the $\d$-functions defines
the restriction on the shell. The both restrictions
$$
P=\sum^m_{k=1}q_k,~P=\sum^n_{k=1}p_k
$$
are compatible since the amplitude $a_{nm}$ is translationally
invariant. The integration over $P$ gives energy-momentum
conservation law.

Notice now that $\R(\a,z)$ is defined through the generating
functional
$$
\R_0(\vp)=Z(\vp_+)Z^* (-\vp_-)=
$$\be
=\int Du_+Du_-
e^{iS_+(u_+)-iS_-(u_-)} e^{-iV_+(u_++\vp_+)+iV_-(u_--\vp_-)}.
\l{2.13a}\ee
Then, we can consider the `closed-path' boundary condition:
\be
\int_{\s_{\infty}}d\sigma_{\mu}u_+\pa^{\mu}u_+=
\int_{\s_{\infty}}d\sigma_{\mu}u_-\pa^{\mu}u_-,
\l{2.13b}\ee
instead of (\r{2.6},\r{2.6a}). The natural solution of this boundary
condition is:
\be
u_+(x\in\s_\infty)=u_-(x\in\s_\infty)=u(x\in\s_\infty).
\l{2.13c}\ee
It provides cancelation of the surface term on the remote
hypersurface $\s_\infty$ independently on the `value' of the field
$u(x\in\s_\infty)$.

Considering the system with the large number of particles, we
can simplify calculations choosing the CM frame $P=(P_0 =E,\vec
0)$.  It is useful also $^{23}$ to rotate the contours of
integration over
$$
\a_{0,k}:~\a_{0,k}=-i\b_k, \Im\b_k =0, k=1,2.
$$
Then $\R(\b,z)$ have a meaning of the density matrix, where $\b$ would
have, in the some definite case $^{24}$, meaning of the inverse
temperature and $z$ is the activity $^{25}$.


It was shown in $^{14}$ that the unitarity condition unambiguously
determines contributions in the  path integrals for $\R$. Exist the
statement:

{\it S1. The density matrix $\R(\a,z)$ has following representation}:
\be
\R(\a,z)= e^{-i\K(je)}\int
DM(u)e^{iS_O(u)-iU(u,e)}e^{N(\a,z;u)}\equiv{\cal O}(u)e^{N(\a,z;u)}.
\l{2.14}\ee
It should be underlined that this representation is strict and is
valid for arbitrary Lagrange theory of arbitrary dimensions. The
derivation of (\r{2.14}) is given in Appendix A.

Expansion over the operator:
\be
\K(je)=\f{1}{2}\Re\int_{C_+} dx dt \f{\d}{\d j(x,t)}\f{\d}{\d
e(x,t)}
\equiv \f{1}{2}\Re\int_{C_+} dx dt \hat{j}(x,t)\hat{e}(x,t)
\l{2.15}\ee
generates the perturbation series. We will assume that
this series exist (at least in Borel sense). The variational
derivatives in (\r{2.15}) are defined as follows:
$$
\f{\d \phi(x,t\in C_i)}{\d \phi(x',t'\in C_j)}=\d_{ij}\d
(x-x') \d (t-t'),~~~ i,j=+,-
$$
where $C_i$ is the Mills time contour. The auxiliary variables
$(j,e)$ must be taken equal to zero at the very end of calculations.

The functionals $U(u,e)$ and $S_O(u)$ are defined
by the equalities:
\be
S_O(u)=(S(u+e)-S(u-e))
+2\Re\int_{C_+}dx dte(x,t)(\pa^2+m^2)u(x,t),
\ee
\be
U(u,e)= V(u+e)-V(u-e)-2\Re\int_{C_+} dx dte(x,t)v'(u),
\l{2.16}\ee
where $S(u)$ is the free part of the Lagrangian and $V(u)$
describes interactions. The phase $S_O(u)$ is not equal to zero
if $u$ have the nontrivial topological charge $^{14}$. We will
discusses carefully this question later.

The measure $DM(u,p)$ has the form:
\be
DM(u)=\prod_{x,t} du(x,t)
\d \le(\f{\d(S(u)-V(u))}{\d u(x,t)}+j(x,t)\ri).
\l{2.17}\ee
The functional $\d$-function in the measure means that the necessary
and sufficient set of contributions in the integral over $u(x,t)$ is
defined by the classical equation:
\be
-\f{\d(S(u)-V(u))}{\d u(x,t)}=j(x,t),
\l{2.17a}\ee
 `disturbed' by the quantum source $j(x,t)$

For further calculation another representation will be useful. If we
insert into the integral (\r{2.14})
$$
1=\int\prod_{x,t}dp(x,t)\d(p(x,t)-\dot{u}(x,t))
$$
then the measure $DM$ takes the form:
$$
DM(u,p)=\prod_{x,t} du(x,t)dp(x,t)\t
$$\be\t
\d\le(\dot{u}(x,t)-\frac{\d H_j (u,p)}{\d p(x,t)}\ri)
\d\le(\dot{p}(x,t)+\frac{\d H_j (u,p)}{\d u(x,t)}\ri)
\l{2.17b}\ee
with the total Hamiltonian
\be
H_j(u,p)=\int dx \le\{ \f{1}{2}p^2 +\f{1}{2}(\nabla u)^2 +
v(u)-ju \ri\}.
\l{2.18}\ee
Last one includes the energy $ju$ of quantum fluctuations. The
measure (\r{2.17b}) describes motion in the symplectic space
$(u,p)\in V$.  But it should be underlined that used expansion is
not the Lagrange transformation. So, generally, it is quite possible,
considering $x$ as the index of space sell, that not all of $p(x,t)$
are the independent variables. For this reason the measure
(\r{2.17b}) has mostly a Lagrange meaning.

The measure (\r{2.17b}) contains following information $^{10,14}$:

{\bf a.} {\it Only the $strict$ solutions of equations
\be
\dot{u}-\frac{\d H_j (u,p)}{\d p}=0,~
\dot{p}+\frac{\d H_j (u,p)}{\d u}=0
\l{2.19}\ee
at $j=0$ should be taken into account}. This `rigidness' means
absence in the formalism of the pseudo-solution (similar to
multi-instanton, or multi-kink) contributions;

{\bf b.} {\it $\R(\a,z)$ is described by the $sum$ of all solutions of
eq.(\r{2.19})}, independently from theirs `nearness' in the functional
space;

{\bf c.} {\it The field disturbed by $j(x)$ belongs to the same
manifold (topology class) as the classical field defined by
(\r{2.19})} $^{10}$.

{\bf d.} {\it The consequence of properties {\bf b.} and {\bf c.} is
the selection rule: quantum dynamics is realized in the coset space
of highest dimension} $^{10}$. This, excluding from consideration
the pure `meson' sector.

The particles density
\be
N(\a,z;u)=N_+(\a_+,z_+;u)+N_-(\a_-,z_-;u),
\l{2.20}\ee
where
\be
N_\pm(\a_\pm,z_\pm;u)=\int d\O_1(q)e^{i\a_\pm
q}z_\pm(q)\le|\Ga(q;u)\ri|^2,
\l{2.21}\ee

The `vertex' $\Ga(q;u)$ is the function of the external particles
momentum $q$ and is the linear functional of $u(x)$:
\be
\Ga(q;u)=-\int dx e^{iqx} \f{\d S(u)}{\d u(x)}=
\int dx e^{iqx}(\pa^2 +m^2)u(x) ,~~q^2=m^2,
\l{2.22}\ee
for the mass $m$ field. This parameter presents the momentum
distribution of the interacting field $u(x)$ on the remote
hypersurface $\s_\infty$ $if$ $u(x)$ is the regular function. Notice,
the operator cancels the mass-shell states of $u(x)$.

Generally $\Ga(q;u)$ is connected directly with $external$ particles
properties and sensitive to the symmetry of the interacting fields
system $^{26}$.

The construction (\r{2.22}) means, because of the operator $(\pa^2
+m^2)$ and remembering that the external states should be mass-shell
by definition $^{9}$, the solution $\R(\a,z)=0$ is actually possible
for particular topology (compactness and analytic properties) of
$quantum$ field $u(x)$. So, $\Ga(q;u)$ carry following remarkable
properties: (i) it directly defines the observables, (ii) is defined
by the topology of $u(x)$. Notice that the space-time topology of
$u(x,t)$ becomes important calculating integral (\r{2.22}) by parts.
This procedure is available if $u(x,t)$ is the regular function. But
the $quantum$ fields are always singular. Therefore, the solution
$\Ga(q;u)=0$ is valid iff the semiclassical approximation is exact,
i.e. the particle production is the pure quantum effect.  Just this
situation is realized in the soliton sector of sin-Gordon model.


Let $G$ be the symmetry of the problem and let $G_c$ be the
symmetry of the  solution $u_c$. Then

{\it S2. The measure (\r{2.17b}) admits the transformation:
\be
u_c:~(u,p)\to (\x,\eta)\in W=G/G_c.
\l{2.23}\ee
and transformed measure has the form:
\be
DM(u,p)=
\prod_{x,t\it C}d\x(t) d\eta(t)
\d \le(\dot{\x}-\frac{\d h_j (\x,\eta)}{\d\eta}\ri)
\d \le(\dot{\eta}+\frac{\d h_j (\x,\eta)}{\d\x}\ri),
\l{2.24}\ee
where $h_j (\x,\eta)=H_j(u_c,p_c)$ is the transformed Hamiltonian.
\be
h_j(\x ,\eta;t) =h(\eta) -\int dx j(x,t) u_c(x;\x ,\eta )
\l{2.25}\ee
and $u_c(x;\x ,\eta )$ is the soliton solution parametrized
by $(\x,\eta )$.}

The proof of eq.(\r{2.24}) is the same as for the Coulomb problem
considered in $^{14}$. But the case of the $(1+1)$-dimensional model
needs the additional explanations. First of all, one must introduce
the functional
\be
\D(u,p)=\int \prod_t d^N \x (t) d^N \eta (t)
\prod_{x,t}\d (u(x,t)-u_c(x;\x ,\eta))
\d (p(x,t)-p_c(x;\x ,\eta)),
\l{2.26}\ee
The equalities
\be
u(x,t)=u_c(x;\x ,\eta),~~p(x,t)=p_c(x;\x ,\eta)
\l{2.27a}\ee
assume that for given $u(x,t)$ and $p(x,t)$ one can hide the $t$
dependence into the $N$ functions $\x=\x(t)$ and $\eta=\eta(t)$. It is
assumed that this procedure can be done for arbitrary $x$. In other
respects functions $u(x,t)$ and $p(x,t)$, and therefore,
$u_c(x;\x,\eta)$ and $p(x;\x,\eta)$, are arbitrary.

For more confidence, one may divide the space onto the $N$ cells and
to each $(u,p)_x$ we may adjust $(\x,\eta)_x$. Quiet possible that
$(\x,\eta)$ are $x$ independent. In this degenerate case $\D\sim
(\d(0))^k$, where $k\leq N$ is the degree of the degeneracy. We will
omit the index $x$ considering $(\x,\eta)_x$ as the vector of the
necessary dimension.

If $(\x,\eta)$ are the solutions of (\r{2.27a}), then
\be
\D(u,p)=\int \prod_t d\x'(t) d\eta'(t)
\d(u_c^\x\x'+u_c^\eta\eta')\d(p_c^\x\x'+p_c^\eta\eta')=
\D_c(\x,\eta)\neq0,
\l{2.27}\ee
where, for instance, $u_c^X=\pa u_c(x;\x,\eta)/\pa X$, $X=\x,\eta$.
Notice importance of last condition. If it fulfilled then one may
insert into (\r{2.14}), with measure (\r{2.24}),
\be
1=\f{\D(u,p)}{\D_c(\x,\eta)}
\l{2.28}\ee
and integrate over $u(x,t)$ and $p(x,t)$. Notice that the possible
infinite factor $(\d(0))^k$ would be canceled in the ratio (\r{2.28}).

The Jacobian of transformation
\newpage
$$
J=\int \f{DuDp}{\D_c(\x,\eta)}\prod_{x,t}
\d \le(\dot{u}-\frac{\d H_j (u,p)}{\d p}\ri)
\d \le(\dot{p}+\frac{\d H_j (u,p)}{\d u}\ri)
$$\be
\t\d (u(x,t)-u_c(x;\x ,\eta))\d (p(x,t)-p_c(x;\x ,\eta)),
\l{2.28a}\ee
is proportional to functional $\d$-functions again. To have the
transformation, we should use the last two $\d$-functions. Notice, if
the first two $\d$-functions are used to calculate $J$, then last two
$\d$-functions realize the constraints. In result,
\be
J=\f{1}{\D_c(\x,\eta)}\prod_{x,t}
\d\le(\dot{u}_c-\frac{\d H_j (u_c,p_c)}{\d p_c}\ri)
\d\le(\dot{p}_c+\frac{\d H_j (u_c,p_c)}{\d u_c}\ri)
\l{2.29}\ee
It should be underlined that $u_c$ and $p_c$ are $arbitrary$
functions of $\x$ and $\eta$, i.e. on this stage we make the
transformation of arbitrary functions $u(x,t)$ and $p(x,t)$ on the
new arbitrary functions $u_c(x;\x,\eta)$ and $p_c(x;\x,\eta)$,
where, generally speaking, $\x=\x(x,t)$ and $\eta=\eta(x,t)$. Then
$\D_c$ is the corresponding determinant.

The expression (\r{2.29}) can be rewritten identically to the form:
$$
J= \f{1}{\D_c(\x,\eta)}
\int\prod_{x,t} d\x'(t) d\eta'(t)\t
$$$$\t
\d\le(\x'-\le(\dot{\x}-\frac{\d h_j(\x,\eta;t)}{\d\eta}\ri)\ri)
\d\le(\eta'-\le(\dot{\eta}+\frac{\d h_j(\x,\eta;t)}{\d\x}\ri)\ri)\t
$$$$
\t\d\le(u_c^\x\x'+u_c^\eta\eta'+
\{u_c,h_j\}-\f{\d H_j}{\d p_c(x,t)}\ri)\t
$$\be
\t\d\le(p_c^\x\x'+p_c^\eta\eta'-
\{p_c,h_j\}+\f{\d H_j}{\d u_c(x,t)}\ri),
\l{2.30}\ee
where $\{,\}$ is the Poisson bracket.

Let us assume now that the auxiliary function $h_j(\x,\eta;t)$ is
chosen so that the equalities
\be
\{u_c,h_j\}=\f{\d H_j}{\d p_c(x,t)},~~~
\{p_c,h_j\}=-\f{\d H_j}{\d u_c(x,t)}.
\l{2.31}\ee
are satisfied identically. Then, taking into account the condition
(\r{2.27}), one can find:
\be
J=\d\le(\dot{\x}-\frac{\d h_j(\x,\eta;t)}{\d\eta}\ri)
\d\le(\dot{\eta}+\frac{\d h_j(\x,\eta;t)}{\d\x}\ri).
\l{2.32}\ee
This ends the transformation. Notice that the determinant $\D_c$ was
canceled identically.

The transformation specify by the equations (\r{2.31}) the function
$h_j$. It assumes that one can find such functions
$u_c=u_c(x;\x,\eta)$ and $p_c=p_c(x;\x,\eta)$, with property
(\r{2.27}), that (\r{2.31}) has unique solution $h_j(\x,\eta;t)$.

Let us convert the problem assuming that just $h_j$ is known. It is
natural to assume that
\be
h_j(\x,\eta;t)=H_j(u_c,p_c),
\l{2.33}\ee
then $u_c$ and $p_c$ are defined by the equations (\r{2.31}) and
\be
\dot{\x}=\frac{\d h_j(\x,\eta;t)}{\d\eta}~
\dot{\eta}=-\frac{\d h_j(\x,\eta;t)}{\d\x}.
\l{2.34}\ee
It is not hard to see that (\r{2.31}) together with (\r{2.34}) are
equivalent to incident equations (\r{2.19}). This is seen from the
following chain of equalities:
$$
\dot{u}_c(x;\x\eta)=u_c^\x\dot{\x}+u_c^\eta\dot{\eta}=
$$$$=
u_c^\x\frac{\pa h_j(\x,\eta;t)}{\pa\eta}-
u_c^\eta\frac{\pa h_j(\x,\eta;t)}{\pa\x}=\{u_c,h_j\}=
\f{\d H_j}{\d p_c(x,t)}
$$
and the same we have for $p_c$. Therefore $(u_c,p_c)$ is the
classical phase space flow and the space $W_G$, labelled by
$(\x,\eta)$, is the coset space $G/G_c$.

In result, new measure takes the form (\r{2.24}), i.e. $\x$ and
$\eta$ should obey the equations (\r{2.34}):
\be
\dot{\x}= \o (\eta) -\int dx j(x,t) \f {\pa u_N (x;\x,\eta)}{\pa
\eta},~
\dot{\eta}=\int dx j(x,t) \f {\pa u_N (\x, \eta)}{\pa \x},
\l{2.35}\ee
where $\o(\eta)\equiv{\pa h(\eta)}/{\pa \eta}$. Hence the source of
quantum perturbations are proportional to the time-local tangent
vectors
$$
\int dx \pa u_N(x;\x,\eta)/\pa \eta, ~
\int dx \pa u_N (x;\x,\eta)/\pa\x
$$
to the soliton configurations. It suggests the idea $^{14}$ to split
the `Lagrange' sources:
$$
j(x,t) \rar (j_{\x}, j_{\eta})(t).
$$

The mechanism of splitting was described in $^{10}$. Resulting
operator ${\cal O}(u_c)$, defined in (\r{2.14}), has the same
structure.  But new perturbations generating operator
\be
\K(e_{\x},e_{\eta};j_{\x},j_{\eta})
=\f{1}{2}Re\int_{C_+} dt \{\hat{j}_{\x} (t)\cdot \hat{e}_{\x}(t)+
\hat{j}_{\eta}(t)\cdot \hat{e}_{\eta}(t)\}.
\l{2.36}\ee
The measure takes the form:
\be
DM(\x,\eta)=\prod_{t}d\x(t) d\eta(t)
\d (\dot{\x}- \o (\eta) - j_{\x}(t))
\d (\dot{\eta} - j_{\eta}(t))
\l{2.37}\ee
The effective potential $U=U(u_c;e_c)$ with
\be
e_c(x,t)=e_{\x}(t) \cdot \f{\pa u_N (x;\x,\eta)}{\pa \eta (t)}-
e_{\eta}(t) \cdot \f{\pa u_N (x;\x, \eta)}{\pa \x (t)}.
\l{2.38}\ee

Notice that the space degree of freedom is disappeared from our
consideration.

\vskip 0.5cm
{\large\bf III. Multiple production in sin-Gordon model}
\vskip 0.3cm


We would consider the theory with Lagrangian
\be
L=\f{1}{2}(\pa_{\mu}u)^2 + \f{m^2}{\la^2}[\cos(\la u)-1] .
\l{3.1}\ee
It is well known that this field model possess the soliton excitations
in the (1+1) dimension.

Formally nothing prevents to linearize partly our problem
considering the Lagrangian
$$
L=\f{1}{2}[(\pa_{\mu}u)^2 - \a m^2 u^2] +
\f{m^2}{\la^2}[\cos(\la u)-1 + \a\f{\la^2}{2} u^2]\equiv
$$\be
\equiv S(u)-v(u)
\l{3.2}\ee
The last term $v(u)=O(u^4)$ describes interactions. Corresponding
vertex function is
\be
\Ga (q;u)=\int dx dt e^{iqx} (\pa^2 +m^2)u(x,t),~~~q^2 =m^2.
\l{3.3}\ee
It should be noted here that chosen in (\r{3.2}) division onto the
`free' and `interaction' parts did not affects the equation of
motion, see (\r{2.17a}), and effective potential, see (\r{2.16}),
i.e. in this sense $\a$ may be chosen arbitrary.  But $\a$ will arise
in the definition of the mass: one should change $m^2\to\a m^2$ in
(\r{3.3}).  This means that our $S$-matrix approach requires
additional, external, normalization condition for the mass shell.  We
will choose $\a=1$ assuming that $m$ is the measured mass of the
`meson'.

We assume that $u(x,t)$ belongs to Schwarz space:
\be
u(x,t)|_{|x| =\infty}=0~(mod \frac{2\pi}{\la}).
\l{3.4}\ee
This means that $u(x,t)$ tends to zero $(mod \frac{2\pi}{\la})$ at
$|x|\rar \infty$ faster then any power of $1/|x|$.

The $\nu$-soliton classical Hamiltonian $h_\nu$ is the sum:
\be
h_\nu(\eta)=\int dr \s (r)\sqrt{r^2+m^2} +\sum^{\nu}_{i=1}h(\eta_i),
\l{3.5}\ee
where $\sigma (r)$ is the continuous spectrum and $h(\eta)$ is the
soliton energy. Notice absence of the energy of soliton interactions.

The $\nu$-soliton solution $u_\nu$ depends on the $2\nu$ parameters.
Half of them $\nu$ can be considered as the position of solitons and
other $\nu$ as the solitons momentum. Generally, at $|t|\to\infty$ the
$u_\nu$ solution decomposed on the single solitons $u_s$ and on the
double soliton bound states $u_b$:
\be
u_\nu(x,t)=\sum^{n_1}_{j=1}u_{s,j}(x,t)+\sum^{n_2}_{k=1}u_{b,k}(x,t)+
O(e^{-|t|})
\l{3.5a}\ee
For this reason the one soliton $u_s$ and two-soliton bound
state $u_b$ would be the main elements of our formalism.  Its
$(\x,\eta)$ parametrizations, i.e. the solution of eq.(\r{2.31}),
has the form $^{27}$:
\be
u_s(x;\x,\eta)=-\f{4}{\la}\arctan\{\exp(mx\cosh\b\eta
-\x)\},~~~ \b =\f{\la^2}{8}
\l{3.6} \ee
and
\be
u_b(x;\x,\eta)=
-\f{4}{\la}\arctan\le\{\tan\f{\b\eta_2}{2}
\f{mx\sinh \f{\b\eta_1}{2}\cos \f{\b\eta_2}{2}-\x_2}
{mx\cosh \f{\b\eta_1}{2}\sin \f{\b\eta_2}{2}-\x_1}\ri\}.
\l{3.7}\ee

The $(\x,\eta)$ parametrization of solitons individual energies $h(\eta)$
takes the form:
$$
h_s(\eta)=\f{m}{\b}\cosh \b\eta,~~~
h_b(\eta)=\f{2m}{\b}\cosh \f{\b\eta_1}{2}\sin\f{\b\eta_2}{2}\geq 0.
$$
The bound-state energy $h_b$ depends on $\eta_2$ amd $\eta_1$.
First one defines inner motion of two bounded solitons and second one
the bound states center of mass motion. Correspondingly we will call
this parameters as the internal and external ones. Note that the
inner motion is periodic, see (\r{3.7}).


Following to the definition of the Dirac measure one should sum over
all solutions of the Lagrange equation, see the property {\bf b.}
in Sec.2. As follows from the equality:
$$
\sum_{\{u_c\}}=\int_{W_G} d\x_0 d\eta_0 \s(u;\x_0,\eta_0)
$$
we should define the density $\s(u;\x_0,\eta_0)$ of states in the
element of the coset space $W_G$. The Faddeev-Popov $ansatz$ is used
for this purpose $^4$.

In our approach, performing the transformation into the coset space
$W_G$, we define the density $\s(u;\x_0,\eta_0)$. Indeed, using the
definition:
$$
\int Dx\prod_t\d(\dot{x})=\int dx(0)=\int dx_0
$$
the functional integrals with measure (\r{2.37}) are reduced to the
ordinary ones over the initial data $(\x,\eta)_{0}$.

But it is important here to trace on the following question. One can
note that, at first glance, integration over $(\x,\eta)_{0}$ may only
give $\R\sim V_{0}^1$, where $V_{0}$ is the zero modes volume,
i.e. is a volume of the $W_G$ space.  On other hand, as follows from
definition of $\R\sim |a_{nm}|^2$, one may expect that $\R\sim
V_{0}^2$.  This discrepancy should have an explanation.

Remembering definition of $\R$ as the squire of amplitudes, we
should defined the contributions on the whole time contour
$C=C_++C_-$, see (\r{2.5}), to take into account the input condition
that the trajectories $u_+(t\in C_+)$ and $u_-(t\in C_-)$ are
absolutely independent in the frame of the `closed-path' boundary
condition (\r{2.13c}):
\be
u_c(x,t\in \pa C_+)=u_c(x,t\in \pa C_-),
\l{3.8}\ee
where $\pa C_\pm$ is the boundary of $C_\pm$. Other directions to the
$\s_|infty$ are not important here.

Then, if we introduce $(\x,\eta)(t\in C_\pm)|_0\equiv
(\x_0,\eta_0)_\pm$, one should have in mind that, generally speaking,
$(\x_0,\eta_0)_+\neq (\x_0,\eta_0)_-$ and the integration over them
should be performed independently. This may explain above
discrepancy and one should have $\R\sim V_{0}^2$.

It is not hard to see that for our topological solitons the condition
(\r{3.8}) leads to the equalities:
\be
(\x_0,\eta_0)_+=(\x_0,\eta_0)_-=(\x_0,\eta_0).
\l{3.9}\ee
To see this it is enough to inserting (\r{3.6}), or (\r{3.7}), into
(\r{3.8}) and take into account that at $t\in \pa C_\pm$ the
estimation (\r{3.5a}) is right.

Solution (\r{3.9}) means that, for arbitrary functional $F(\x,\eta)$,
\be
\int \prod_{t\in C_++C_-}d\x d\eta \d(\dot\x)\d(\dot\eta)F(\x,\eta)=
\int d\x_{0+}d\eta_{0+}\int d\x_0d\eta_0 F(\x_0,\eta_0).
\l{3.9a}\ee
Therefore, $\R\sim V_0^2$. We will put out the integrals over
inessential variables $\x_{0+}$ and $\eta_{0+}$.

It should be underlined that (\r{3.9}) is the consequence of the
conservation of the topological charge: the solitons by this
reason are the stable formation and, therefore, to satisfy the closed
path boundary condition, one should have (\r{3.9}).


Performing the shifts:
\ba
\x_i (t) \rar \x_i (t) + \int dt' g(t-t') j_{\x,i}(t') \equiv
\x_i (t) +\x'_i (t),
\n \\
\eta_i (t) \rar \eta_i (t) + \int dt' g(t-t') j_{\eta ,i}(t') \equiv
\eta_i (t) +\eta'_i (t),
\n\ea
we can get the Green function $g(t-t')$ into the operator
exponent:
\be
\K(ej) =\f{1}{2}\int dtdt'\Th (t-t')\{\hat{\x}'(t')\cdot
\hat{e}_{\x}(t)+
\hat{\eta}'(t')\cdot \hat{e}_{\eta}(t)\}.
\l{3.10}\ee
since the Green function $g(t-t')$ of the transformed theory is the
step function $^{10}$:
\be
g(t-t')=\Th(t-t')
\l{3.11}\ee
Such Green function allows to shift $C_{\pm}$ on the real-time axis.
This, noting (\r{3.9}), excludes doubling of the degrees of freedom.

Notice the Lorentz noncovariantness of our perturbation theory with
Green function (\r{3.11}).

The measure takes the form:
\be
D^\nu M(\x,\eta)=\prod^{\nu}_{i=1}\prod_{t}d\x_i(t) d\eta_i(t)
\d (\dot{\x}_i- \o (\eta+\eta'))\d (\dot{\eta}_i).
\l{3.12}\ee
The interactions are described by
\be
U(u_\nu;e_c)=-\f{2m^2}{\la^2}\int dx dt \sin\la u_\nu~
(\sin \la e_c -\la e_c)
\l{3.14}\ee
with
\be
u_\nu=u_\nu (x;\x+\x',\eta+\eta')
\ee
and $e_c$ was defined in (\r{2.38}).

The equations:
\be
\dot{\x}_i=\o (\eta_i+\eta'_i)
\ee
are trivially integrable. In quantum case $\eta'_i \neq 0$ this
equation describes motion in the nonhomogeneous and anisotropic
manifold. So, the expansion over $(\hat{\x'},~\hat{e}_{\x},~\hat{\eta}',
~\hat{e}_{\eta})$ generates the local in time fluctuations of
$W_G$ manifold. The weight of this fluctuations is defined by
$U(u_\nu;e_c)$.

Expansion of $\exp\{\K(je)\}$ gives the `strong coupling'
perturbation series. The analyses shows that $^{14}$

{\it S3.Action of the integro- differential operator $\h{\cal O}$
leads to following representation}:
\be
\R(\a,z)=\int_{W_G}\le\{
d\x(o)\cdot\f{\pa}{\pa\x(0)}R^\x(\a,z)+
d\eta(0)\cdot\f{\pa}{\pa\eta(0)}R^\eta(\a,z)\ri\}.
\l{3.13}\ee
This means that the contributions into $\R$ are accumulated
strictly on the boundary `bifurcation manifold' $\pa W_G$. The prove
of this important result was given in $^{10,14}$ and we will use it
without comments.


We would divide calculations on two parts. First of all, we would
consider the semiclassical approximation and then we will show that
this approximation is exact.

Performing the last integration we find:
\be
\R(\a,z)=\int \prod^{\nu}_{i=1} \{d\x_0 d\eta_0\}_i
e^{-i \hat{K}}e^{iS_O(u_\nu)}e^{-iU(u_\nu;e_c)}
e^{N(\a,z;u_\nu)}
\l{5.1}\ee
where
\be
u_\nu=u_\nu (\eta_0 +\eta',\x_0 + \o (t) +\x').
\l{5.2}\ee
and
\be
\o(t)=\int dt'\Th (t-t')\o(\eta_0 +\eta')(t')
\l{5.3}\ee

In the semiclassical approximation $\x'=\eta'=0$ we have:
\be
u_\nu=u_\nu (x;\eta_0 ,\x_0 + \o(\eta_0)t).
\l{5.7}\ee
Notice that the surface term
\be
\int dx^\mu\pa_{\mu}(e^{iqx}u_\nu)=0.
\l{5.8}\ee
Then
\be
\int d^2x e^{iqx}(\pa^2 +m^2)u_\nu (x,t)
=-(q^2-m^2)\int d^2x e^{iqx}u_\nu (x,t) =0
\l{5.9}\ee
since $q^2$  belongs to the mass shell by definition. The condition
(\r{5.8}) is satisfied for all $q_{\mu}\neq0$ since $u_\nu$ belong
to the Schwarz space. Therefore, in the semiclassical approximation
$R^c(\a,z)$ is the trivial function of $z$: $\pa R^c(\a,z)/\pa z=0$.

Expending the operator exponent in (\ref{5.1}), we find that action
of the operators $\hat{\x}'$, $\hat{\eta}'$ create the terms
\be
\sim\int d^2x e^{iqx} \th (t-t') (\pa^2 +m^2)u_\nu (x,t) \neq 0.
\l{5.12}\ee
So, generally $R(\a,z)$ is the nontrivial function of $z$.

Now we will show that the semiclassical approximation is exact in
the soliton sector of the sin-Gordon model. The structure of the
perturbation theory is readily seen in the `normal- product' form:
\be
R(\a,z)=\sum_\nu \int \prod^{N}_{i=1} \{d\x_0 d\eta_0\}_i
\n\\\times
:e^{-iU(u_\nu;\hat{j}/2i)}e^{iS_O(u_\nu)}e^{N(\a,z;u_\nu)}:,
\l{6.1}\ee
where
\be
\hat{j}=\hat{j}_{\x}\cdot\f{\pa u_\nu}{\pa \eta}- \hat{j}_{\eta}\cdot
\f{\pa u_\nu}{\pa \x}=\O \hat{j}_{X}\f{\pa u_\nu}{\pa X}
\l{6.2}\ee
and
\be
\hat{j}_{X}=\int dt' \Th(t-t')\hat{X}(t')
\l{6.3}\ee
with the $2N$-dimensional vector $X=(\x ,\eta)$. In eq.(\r{6.2}) $\O$
is the ordinary symplectic matrix.

The colons in (\r{6.1}) mean that the operator $\hat{j}$ should stay
to the left of all functions. The structure (\r{6.2}) shows that
each order over $\hat{j}_{X_i}$ is proportional at least to the first
order derivative of $u_\nu$ over conjugate to $X_i$ variable.

The expansion of (\r{6.1}) over $\hat{j}_{X}$ can be written using
in the form:
\be
\R(\a,z)=\sum_\nu \int \prod^{\nu}_{i=1} \{d\x_0 d\eta_0\}_i
\le\{\sum^{2\nu}_{i=1}\f{\pa}{\pa X_{0i}}P_{X_i}(u_\nu)\ri\},
\l{6.4}\ee
where $P_{X_i}(u_\nu)$ is the infinite sum of the `time-ordered'
polynomial over $u_\nu$ and its derivatives $^{14}$. The explicit form
of $P_{X_i}(u_\nu)$ is unimportant, it is enough to know, see
(\r{6.2}), that
\be
P_{X_i}(u_\nu)\sim \O_{ij} \f{\pa u_\nu}{\pa
X_{0j}}.
\l{6.5}\ee

Therefore,
\be
\f{\pa}{\pa z}R(\a,z)=0
\l{6.6}\ee
since (i) each term in (\r{6.4}) is the total derivative, (ii) we have
(\r{6.5}) and (iii) $u_\nu$ belongs to Schwarz space.

\vskip 0.5cm
{\large\bf IV. Conclusion}
\vskip 0.3cm

We would like to conclude this paper noting the role of the coset
space $G/G_c$ topology. It was shown that if

(a) $W_G=G/G_c\neq\emptyset$,

(b) $W_G=T^*V$ is the simplectic manifold,

(c) $\pa u_c$ is the flow,

(d) $\pa u_c\bigcap\pa W_G=\emptyset,$\\
then the semiclassical approximation is exact.

For this reason, being absolutely stable, `topological solitons' are
unable to describe the multiple production processes. This property
of the exactly integrable models was formulated also as the absence
of stochastization in the integrable systems $^{28}$. The $O(4)\t
O(2)$-invariant solution of $O(4,2)$-invariant theories $^{29}$ is
noticeably more interesting from this point of view $^{30}$.

\vspace{0.4in}
{\Large \bf Acknowledgement}

We acknowledged to members of the seminar `Symmetries and integrable
systems' of the N.N.Bogolyubov Laboratory of theoretical physics
(JINR) for important discussions. We would like to thank
V.G.Kadyshev\-ski for fruitful interest to described technique and
underling idea.  One of us (JM) was partly granted by Georgian Acad.
of Sciences.

\renewcommand{\theequation}{A.\arabic{equation}}
\appendix\section{Appendix. Derivation of eq.({29})}\0

The generating functional (\r{2.13}) can be written in the form:
\be
\R(\b,z)=e^{-\N(\b,z;\vp)}\R_0(\vp),
\l{a1}\ee
where the particles number operator
\be
\N(\b,z;\vp)=\N(\b_+,z_+;\vp)+\N^*(\b_-,z_-;\vp),
\l{a2}\ee
and
\be
\N(\b_+,z_+;\vp_+)=\int d\O_1(q)\h{\vp}_+^*(q)\h{\vp}_-(q)
e^{-\b_+\e(q)}z_+(q)
\l{a3}\ee
is the produced particles number operator.

The functional $\R_0$ was introduced in (\r{2.13a}):
$$
\R_0(\vp)=Z(\vp_+)Z^* (-\vp_-)=
$$\be
=\int Du_+Du_-
e^{iS_+(u_+)-iS_-(u_-)} e^{-iV_+(u_++\vp_+)+iV_-(u_--\vp_-)}.
\l{a4}\ee
So, the integration over $u_+$ and $u_-$ is not performed
independently: one should take into account the boundary condition
(\r{2.13c}).  We can perform in this integrals the linear
transformation:
\be
u_\pm(x)=u(x)\pm\p(x).
\l{a5}\ee
Then the boundary condition (\r{2.13c}) leads to equality:
\be
\p(x\in\s_\infty)=0,
\l{a5a}\ee
leaving $u(x\in\s_\infty)$ arbitrary.  Last one means that the
integration over this `turning-point' field $u(x\in\s_\infty)$ should
be performed, see Sec.3.

Let us extract in the exponents (\r{a4}) the linear over $(\phi
+\vp)$ term:
$$
V_+(u+(\p+\vp)) - V_-(u- (\p+\vp))=
$$\be
+U(u,\p+\vp)
+2\Re\int_{C_+} dx (\phi (x) +\vp (x))v'(u),
\l{a6}\ee
and
\be
S_+(u+ \vp) -S_-(u- \vp)=S_O (u)
 -2i\Re\int_{C_+} dx\vp (x) (\pa_{\mu}^2 +m^2 )u(x).
\l{a7}\ee
where
$$
2\Re\int_{C_+}=\int_{C_+} + \int_{C_-}.
$$
Notice that the generally speaking, $S_O(u)\neq 0$ if the topology of
field $u(x)$ is nontrivial, see Sec.3.

The expansion over $(\p+\vp)$ can be written in the form:
\be
e^{-iU(u,\phi+\vp)}=e^{\f{1}{2i}Re\int_{C_+}dx \hat{j}(x)
\hat{\vp'}(x)}
e^{i2\Re\int_{C_+}dx dt j(x)(\phi(x)+\vp(x))}e^{-iU(u,\vp')},
\l{a8}\ee
where $\h{j}(x),~\h{\vp'}(x)$ are the variational
derivatives. The auxiliary variables $(j,\vp')$ must be taken equal
to zero at the very end of calculations.

In result,
\ba
&\R_0(\phi)=e^{\f{1}{2i}\Re\int_{C_+} dx \hat{j}(x) \hat{\vp}(x)}\int
Du e^{is_0(u)}e^{-iU(u,\vp)}
e^{i2\Re\int_{C_+}dx (j(x)-v'(u)) \phi(x)}\t
\n\\
&\t\prod_{x}\d (\pa_{\mu}^2 u + m^2 u +v'(u) -j),
\l{a9}\ea
where the functional $\d$-function was defined by the equality:
\be
\prod_{x}\d (\pa_{\mu}^2 u + m^2 u +v'(u) -j)
=\int D'\p e^{-2i\Re\int_{C_+}dx (\pa_{\mu}^2 u + m^2 u +v'(u) -
j)\vp(x)},
\l{a10}\ee
where the prime means that $D'\p$ does not includes the integration
over $\p(x\in\s_\infty)$. This condition is not seen in the
functional $\d$-function because of the definition:
$$
\int \prod_x du(x)\d(\pa_\mu u(x))=\int du(x_\mu\in\s_\infty).
$$

The eq.(\r{a9}) can be rewritten in the
equivalent form:
\be
\R_0(\phi)=e^{-i\hat{K}(j,\vp)}\int
DM(u)e^{is_0 (u) -iU(u,\vp)}
e^{i2Re\int_{C_+}dx \phi(x)(\pa_{\mu}^2 + m^2) u(x)}
\l{a11}\ee
because of the $\d$-functional measure:
\be
DM(u)=\prod_{x} du(x) \d (\pa_{\mu}^2 u + m^2 u +v'(u)  -j),
\l{a12}\ee
with
\be
\K(j\vp)=\f{1}{2}\Re\int_{C_+} dx\hat{j}(x)\hat{\vp}(x).
\l{a13}\ee
Notice at the end that the contour $C_+$ in (\r{a13})
can not be shifted on the real time axis since the Green function of
the equation
$$
\pa_{\mu}^2 u + m^2 u +v'(u)=j
$$
is singular on the light cone.

The action of operator ${\bf N}(\b,z;\h\p)$ maps the interacting
fields system on the physical states. Last ones are `marked' by
$z_\pm$ and $\b_\pm$. The operator exponent is the linear
functional over $\p$ and this allows easily find (\r{2.14}).

\end{document}